\documentclass[pdflatex,sn-mathphys-num]{sn-jnl}
\usepackage{graphicx}
\usepackage{multirow}
\usepackage{amsmath,amssymb,amsfonts}
\usepackage{amsthm}
\usepackage{mathrsfs}
\usepackage[title]{appendix}
\usepackage{xcolor}
\usepackage{textcomp}
\usepackage{manyfoot}
\usepackage{booktabs}
\usepackage{algorithm}
\usepackage{algorithmicx}
\usepackage{algpseudocode}
\usepackage{listings}
\usepackage{hyperref}
\usepackage{url}
\usepackage{rotating}

\theoremstyle{thmstyleone}

\theoremstyle{thmstyletwo}

\theoremstyle{thmstylethree}

\begin{document}

\title{Hybrid DiffractGPT-Rietveld Refinement Framework for Automated X-ray Diffraction Analysis}
\author[1,4]{Charles Rhys Campbell}
\author[2]{Justin Ely}
\author[1]{Jaehyung Lee}
\author[3]{Frank M. Abel}
\author[1,2*]{Kamal Choudhary}

\affil[1]{Department of Materials Science and Engineering, Johns Hopkins University, Baltimore, MD 21218, USA}
\affil[2]{Department of Electrical and Computer Engineering, Johns Hopkins University, Baltimore, MD 21218, USA}
\affil[3]{The Volgenau Department of Physics, United States Naval Academy, Annapolis, MD 21402, USA}
\affil[4]{Department of Physics and Astronomy, West Virginia University, Morgantown, WV 26505, USA}

\affil[*]{Corresponding author: kchoudh2@jhu.edu}

\maketitle

\begin{abstract}
XX-ray diffraction (XRD) is fundamental to structural materials characterization, yet
transforming a raw powder pattern into a refined crystal structure still demands
considerable domain expertise. We present AGAPI-XRD, a hybrid framework integrating
DiffractGPT generative structure prediction, database pattern matching against
JARVIS-DFT and COD, and automated Rietveld refinement and ALIGNN-FF relaxation through a unified API at
\url{https://atomgpt.org/xrd}. First, we used the AGAPI-XRD pipeline to evaluate the crystal structure of a variety of minerals in the RRUFF database that were experimentally characterized using powder x-ray diffraction. Next, we benchmarked the lattice parameter prediction fidelity of the AGAPI-XRD pipeline using a subset of the Alexandria PBE-hull dataset and the subset of RRUFF minerals that have known lattice parameters. AGAPI-XRD returns valid lattice parameters for 79.7\% of the RRUFF benchmark minerals and for 94.8--98.1\% of the Alexandria subset, while identifying a candidate structure for 93.8\% of RRUFF minerals. For this benchmark, pattern matching delivers
the highest accuracy for known phases, while DiffractGPT extends structure generation to
complex materials absent from existing databases. Together, AGAPI-XRD advances
accessible, end-to-end automated crystal structure determination from powder XRD data.
\end{abstract}

\keywords{X-ray diffraction, Rietveld refinement, generative AI, crystal structure determination, materials informatics}

%% =========================================================================
%%  INTRODUCTION
%% =========================================================================
\section{Introduction}\label{sec:intro}

X-ray diffraction (XRD) is the cornerstone of crystallographic characterization as it provides direct access to the periodic arrangement of atoms in solids~\cite{cullity2001, warren1990}. Since the pioneering work of Bragg and the development of powder diffraction methods, XRD has been indispensable for phase identification, lattice parameter determination, and full structural refinement across virtually every class of crystalline material~\cite{rietveld1969, young1993}. The introduction of Rietveld refinement in 1969 transformed powder diffraction from a qualitative fingerprinting technique into a quantitative structural tool, enabling the extraction of atomic positions, thermal parameters, and site occupancies from polycrystalline data~\cite{rietveld1969}.

Despite these advances, the conventional workflow from experimental diffraction pattern to refined crystal structure remains labor-intensive. A typical analysis requires: (i)~phase identification through manual comparison with reference databases such as the Powder Diffraction File (PDF)~\cite{gates2019}, the Crystallography Open Database (COD)~\cite{grazulis2012}, or the Inorganic Crystal Structure Database (ICSD)~\cite{bergerhoff1987}; (ii)~selection of an appropriate structural model as a starting point; (iii)~iterative Rietveld refinement with careful sequential release of parameters~\cite{toby2013}; and (iv)~validation of the refined structure against chemical and physical constraints. Each step demands significant domain expertise, and the quality of the final result depends critically on the initial structural guess, and a poor starting model can lead to convergence at physically meaningless local minima or outright divergence of the refinement~\cite{mccusker1999}.

The advent of machine learning (ML) in materials science has begun to transform many aspects of computational materials research~\cite{choudhary2022review, butler2018, schmidt2019}. Graph neural networks such as ALIGNN and ALIGNN-FF~\cite{alignn2021,choudhary2023unified} have achieved remarkable accuracy in predicting material properties from crystal structures and optimizing atomic structure geometries. Generative models including variational autoencoders~\cite{court2020}, generative adversarial networks~\cite{kim2020gan}, and diffusion models~\cite{xie2022cdvae} have demonstrated the ability to design new crystal structures with targeted properties. More recently, large language models have been applied to materials science through frameworks such as AtomGPT~\cite{atomgpt2024,diffractgpt2025,choudhary2025microscopygpt}, which uses instruction-tuned transformers to predict material properties and generate crystal structures from text descriptions.

Several groups have applied ML specifically to XRD analysis. Park et al.\ developed convolutional neural networks for crystal system classification from powder patterns~\cite{park2017}. Lee et al.\ used deep learning for automated phase identification~\cite{lee2020xrd}. Vecsei et al.\ demonstrated neural network-based lattice parameter extraction~\cite{vecsei2019}. However, these approaches typically address individual components of the analysis pipeline rather than providing end-to-end structure determination. DiffractGPT~\cite{diffractgpt2025} introduced a fundamentally different approach: a generative pretrained transformer that directly predicts complete atomic structures (including lattice parameters, space group, and atomic coordinates) from experimental XRD peak positions and chemical composition. Trained on simulated XRD patterns from the JARVIS-DFT database~\cite{jarvis2020}, DiffractGPT performs rapid inverse design, generating candidate structures in seconds rather than the hours or days required by traditional \textit{ab initio} structure solution methods.

In parallel, the JARVIS (Joint Automated Repository for Various Integrated Simulations) ecosystem~\cite{jarvis2020,choudhary2025jarvis} has established one of the largest open-access computational materials databases, comprising over one million density functional theory (DFT) calculations with consistently computed properties. The JARVIS infrastructure, including databases, tools, and machine learning models, is accessible through the AGAPI (AtomGPT.org API) platform~\cite{agapi2025} at \url{https://atomgpt.org}, which provides programmatic access to the full suite of materials informatics capabilities.

Here we present AGAPI-XRD, a hybrid framework that combines three complementary approaches into a single automated pipeline: (1)~pattern matching against the combined JARVIS-DFT (${\sim}76{,}000$ entries) and COD (${\sim}431{,}000$ entries) structure databases for rapid identification of known phases; (2)~DiffractGPT generative structure prediction for novel phases absent from existing databases; (3)~classical Rietveld refinement (via GSAS-II~\cite{toby2013} or BGMN~\cite{bgmn1998}) for quantitative structural optimization; (4) optional ALIGNN-FF structure optimization \cite{choudhary2023unified}; (5) ALIGNN \cite{alignn2021}/SlaKoNet property predictions \cite{slakonet2024}. We benchmark this pipeline on 276 crystals from the RRUFF powder XRD database~\cite{rruff2015} and on a 1{,}000-structure representative subset of the Alexandria PBE-hull simulated-XRD benchmark~\cite{alexandria} spanning six pipeline configurations, providing a systematic evaluation of combined AI, database-driven, and XRD structure identification against both experimentally characterized natural minerals and high-throughput DFT reference structures.

%% =========================================================================
%%  RESULTS & DISCUSSION
%% =========================================================================
\section{Results \& Discussion}\label{sec:results}
\begin{figure}[t]
\centering
\includegraphics[width=\linewidth]{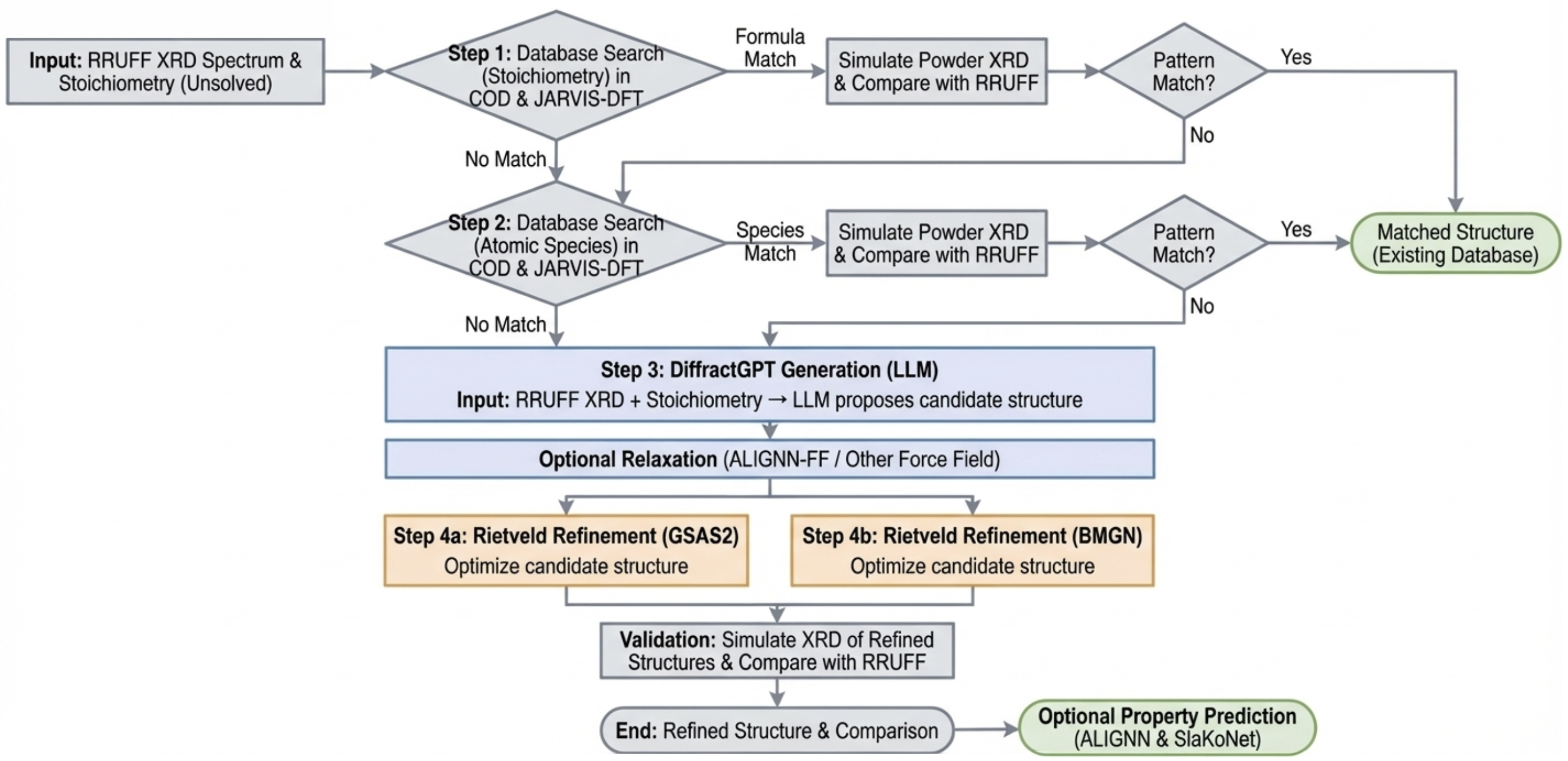}
\caption{\textbf{Schematic of the AGAPI-XRD pipeline.} An experimental powder XRD pattern and chemical formula/elements are submitted to the AtomGPT.org API, which executes a multi-stage workflow. \textbf{Stage~1:} Pattern matching against the combined JARVIS-DFT (${\sim}76{,}000$ entries) and COD (${\sim}431{,}000$ entries) databases using cosine similarity on common-binned intensity profiles. Formula-based search is attempted first; if unsuccessful, a fallback element-based compositional search is triggered. \textbf{Stage~2:} If the pattern is not matched, DiffractGPT generative structure prediction using a fine-tuned Mistral-7B generative pretrained transformer, which takes extracted peak positions and the chemical formula as input and generates complete lattice parameters, space group, and atomic coordinates. \textbf{Stage~3:} Optional structural relaxation with ALIGNN-FF, or another ML force field, as well as an automated Rietveld refinement via GSAS-II or BGMN with sequential cumulative parameter optimization. \textbf{Stage~4:} Optional ML property predictions via ALIGNN (formation energy, band gap, elastic moduli, superconducting $T_c$) and SlaKoNet (tight-binding band structure). The pipeline returns the best-matching crystal structure as a POSCAR file in conjunction with full metadata, similarity scores, refinement statistics, and predicted properties.}
\label{fig:schematic}
\end{figure}

\begin{figure}[t]
\centering
\includegraphics[width=0.7\linewidth]{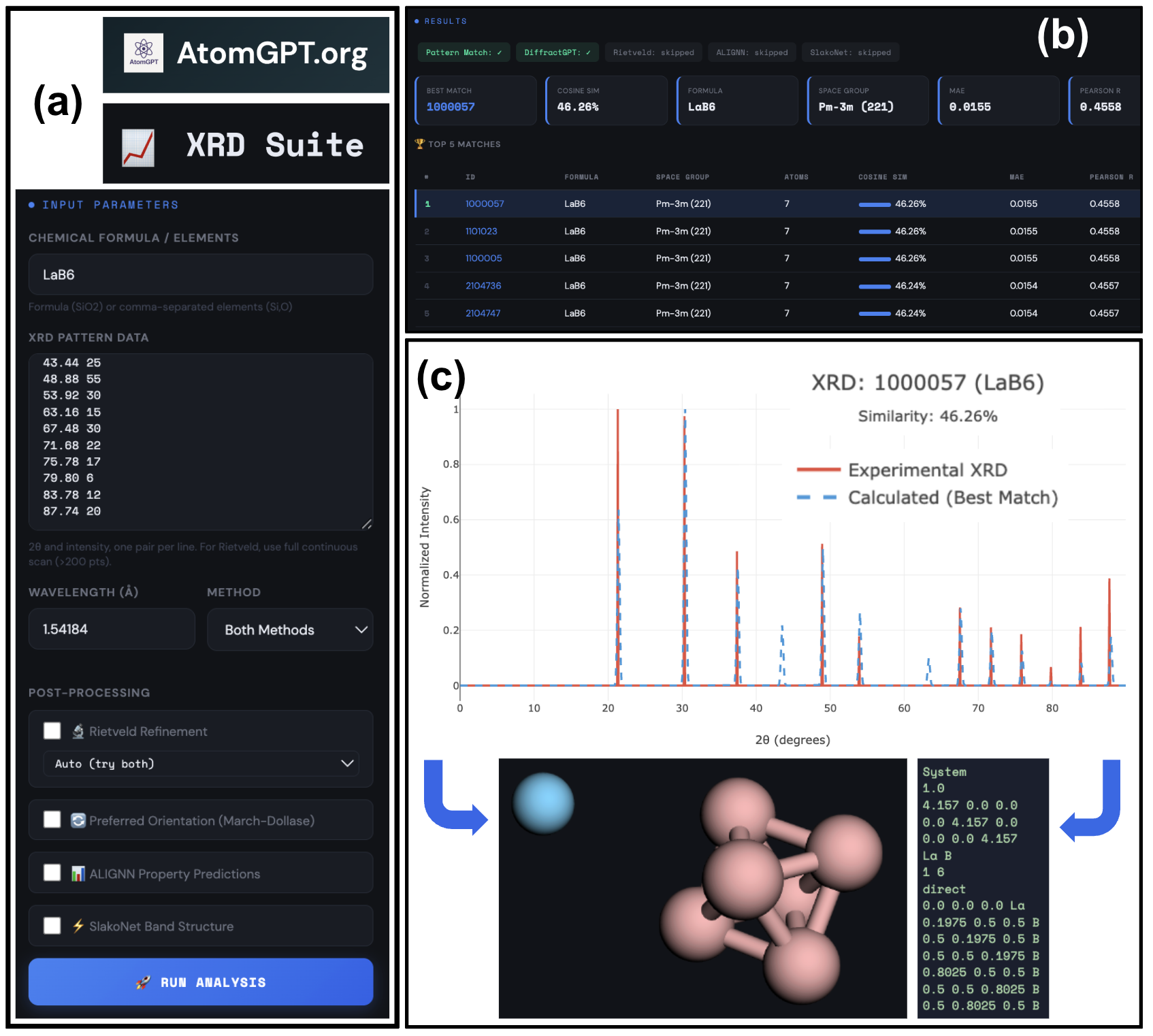}
\caption{\textbf{Web-based AGAPI-XRD interface on AtomGPT.org for automated crystal structure identification from powder XRD data.} \textbf{(a)} Input panel showing submission of an experimental XRD spectrum with a chemical formula or elemental composition, along with user-selectable options for pattern matching, DiffractGPT-based structure generation, Rietveld refinement, and downstream materials-property analysis. \textbf{(b)} Ranked candidate structures returned by the pipeline with match scores and key structural descriptors. \textbf{(c)} Visualization panel displaying the selected crystal structure, its structural representation, and the overlay between input and predicted diffraction patterns. The figure highlights the integrated, end-to-end design of AGAPI-XRD for accessible automated XRD analysis.}
\label{fig:atomgpt_dot_ord_xrd}
\end{figure}

The AGAPI-XRD pipeline is implemented as a RESTful API endpoint (\texttt{/xrd/analyze\_with\_refinement}) within the AtomGPT.org platform (Figure~\ref{fig:schematic}), and is also accessible through a web-based graphical interface (Figure~\ref{fig:atomgpt_dot_ord_xrd}). For a given XRD spectrum and corresponding stoichiometry, the pipeline executes the multi-stage workflow described below. Full mathematical definitions and implementation details for each stage are provided in the Methods (Section~\ref{sec:methods}).

The first stage of the AGAPI-XRD pipeline is pattern-matching. The experimental XRD pattern and chemical formula are compared against the combined JARVIS-DFT and COD databases. For each candidate structure, a theoretical powder XRD pattern is simulated using the kinematic diffraction approximation implemented in the JARVIS analysis module~\cite{jarvis2020}. The similarity between experimental and calculated patterns is quantified by cosine similarity on re-binned intensity profiles (Eq.~\ref{eq:cosine}). Formula-based search is first attempted using reduced chemical formulas comparison with JARVIS-DFT and COD structures. If no match is found, a fallback element-based search is triggered using comma-separated constituent elements, which queries both databases for structures containing the same elemental combination. If no structure is found in either database that share an XRD spectrum and stoichiometry with the candidate crystal, the pipeline proceeds to Stage 2.

The second stage is DiffractGPT structure generation. In parallel or as a complement to pattern matching, DiffractGPT generates candidate crystal structures directly from the XRD data. Peak positions and relative intensities are extracted from the experimental pattern using the \texttt{scipy.signal.find\_peaks} algorithm~\cite{scipy2020} with a height threshold of 0.05, a prominence of 0.02, and a minimum peak separation of $0.5^\circ$ in $2\theta$ (converted to the corresponding number of data points using the pattern's step size). The top~20 peaks (sorted by intensity) are formatted as a text prompt along with the chemical formula and submitted to the DiffractGPT model, a Mistral-7B architecture~\cite{jiang2023mistral} fine-tuned on JARVIS-DFT structure--XRD pairs~\cite{diffractgpt2025}. The model generates a complete structural description including lattice parameters, angles, and fractional atomic coordinates.

The third stage is optional, and it is ALIGNN-FF structural relaxation. The candidate structure can be further relaxed using ALIGNN-FF to reduce local geometric artifacts and provide a more physically plausible starting point for subsequent Rietveld refinement. This step is particularly useful for AI-generated structures, where small errors in atomic positions or lattice geometry can impede refinement convergence.

The fourth stage is Rietveld refinement. The best structure from Stages 1 and 2 serves as the initial model for automated Rietveld refinement. Two refinement engines are supported: GSAS-II~\cite{toby2013} via its scriptable Python interface, and BGMN~\cite{bgmn1998} via the \texttt{dara.refine} package. Refinement proceeds through a sequential cumulative strategy: background (12-term Chebyshev polynomial), unit cell parameters, instrument zero point, profile parameters ($U$, $V$, $W$, $X$, $Y$, SH/L), and sample displacement. An optional March--Dollase preferred orientation correction~\cite{march1932,dollase1986} can be enabled. The weighted profile R-factor ($R_{\mathrm{wp}}$) serves as the primary figure of merit.

The fifth and final stage is optional, and it is material property prediction using surrogate deep learning models within the AtomGPT ecosystem. The identified structure can be passed to ALIGNN~\cite{alignn2021} for rapid ML-based property predictions (formation energy, band gap, elastic moduli, superconducting $T_c$) and to SlaKoNet~\cite{slakonet2024} for tight-binding electronic structure calculations, providing immediate materials informatics context for the identified phase.

\begin{figure}[t]
\centering
\includegraphics[width=0.7\linewidth]{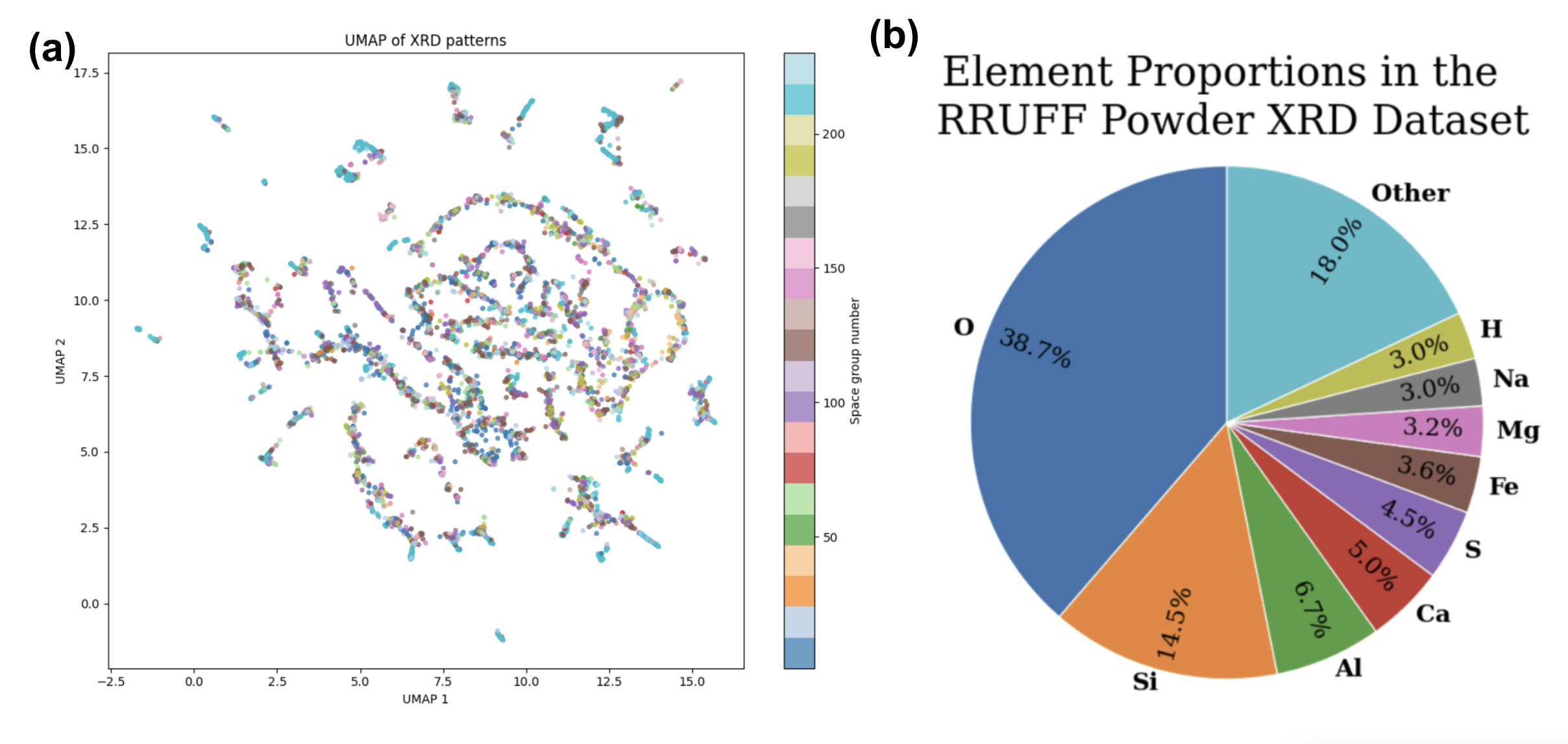}
\caption{\textbf{Visualization of computational and experimental x-ray diffraction dataset.} (a) Converted 76,000 materials in JARVIS-DFT and 431,000 materials in COD database and visualized using UMAP~\cite{mcinnes2018umap} using space group as a color bar. We observed clusters of high-symmetry space groups, especially blue and green space group labels. (b) Elemental composition of the experimental RRUFF database on which the AGAPI-XRD framework will be tested. The top nine most common elements are shown.}
\label{fig:element_freq}
\end{figure}

The structural diversity of the JARVIS-DFT and COD matching databases, together with the elemental composition of the experimental RRUFF benchmark, is summarized in Figure~\ref{fig:element_freq}. The RRUFF Project~\cite{rruff2015} provides high-quality powder XRD data for naturally occurring minerals, collected under standardized conditions with Cu~K$\alpha$ radiation ($\lambda = 1.54184$~\AA). We accessed the RRUFF powder XRD dataset through the JARVIS figshare interface, which provides 1{,}362 entries. After deduplication by mineral name and filtering for valid chemical compositions (rejecting entries with unparseable formulas or HTML-encoded artifacts), 803 unique minerals remained. To focus on structures where lattice parameter comparison is most meaningful and to reduce the impact of supercell ambiguities, we further filtered to minerals with RRUFF-reported lattice parameters $a, b, c \leq 10$~\AA, yielding 276 minerals spanning all seven crystal systems. Of these, 220 minerals additionally had valid predictions for all lattice parameters and form the basis for the lattice-parameter accuracy analysis (Tables~\ref{tab:overall_metrics} and~\ref{tab:crystal_system}), distributed across crystal systems as cubic~(50), orthorhombic~(45), monoclinic~(45), triclinic~(26), hexagonal~(26), tetragonal~(21), and rhombohedral~(7). Each RRUFF entry provides a continuous powder XRD pattern (typically 8{,}501 data points spanning $20^\circ$--$95^\circ$ $2\theta$ at $0.02^\circ$ steps), experimentally determined lattice parameters with uncertainties, the crystal system classification, and the chemical formula with site occupancies.

\begin{figure}[t]
\centering
\includegraphics[width=0.8\linewidth]{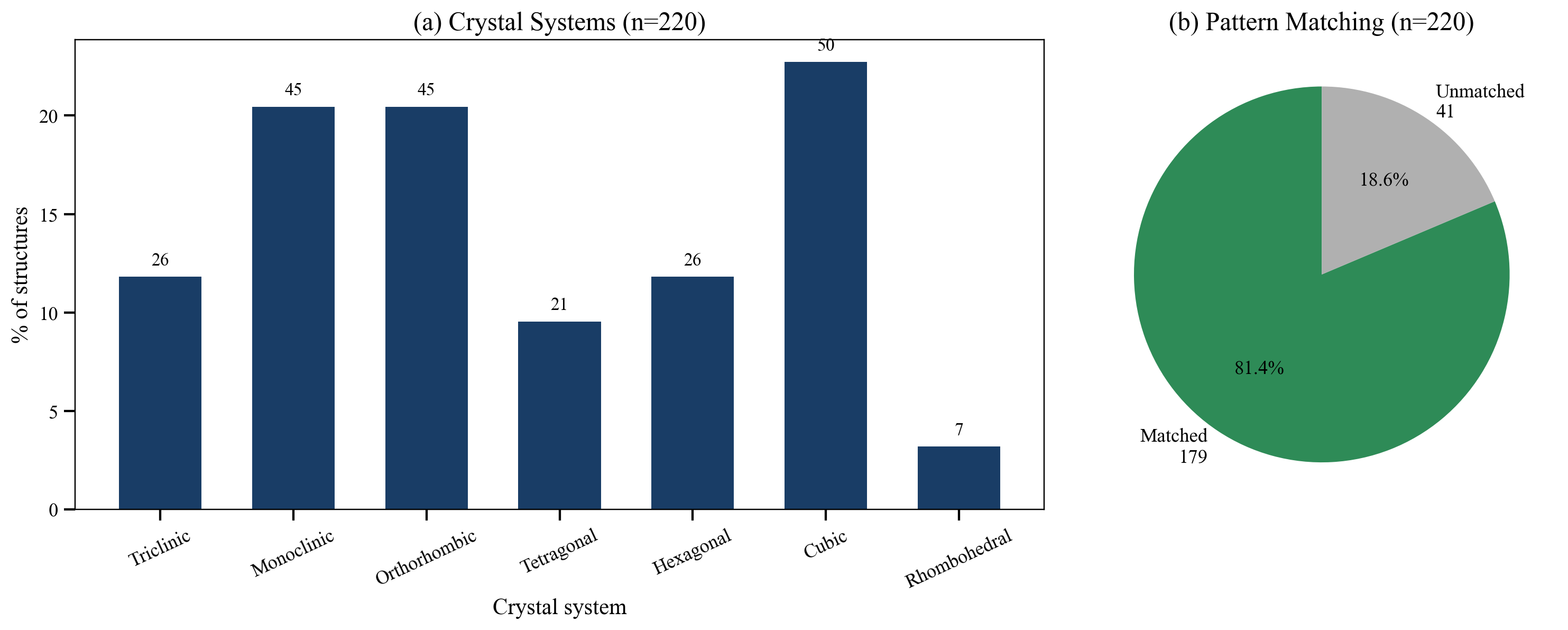}
\caption{\textbf{Aggregate pattern-matching match rate and crystal system histograms of predicted unit cells for three refinement methods.} {Sub-figure (a) shows the crystal system labels for RRUFF database structures after filtering (n=220; see Table~\ref{tab:crystal_system}). Sub-figure (b) shows the percentage of RRUFF structures that were matched and unmatched to existing structures in the JARVIS-DFT and COD databases using the pattern-matching procedure. Approximately 73\% (72.8\%) of structures in the RRUFF database were matched to existing structures.}}
\label{fig:crystal_system_histograms_plus_match_rate_pie}
\end{figure}

To complement this experimental benchmark with a simulated-XRD test set spanning a broader chemical and structural range, we also evaluated AGAPI-XRD on structures from the Alexandria PBE-hull dataset~\cite{alexandria}. This dataset contains approximately 117{,}000 on-hull entries with DFT-relaxed reference lattice parameters. We applied the same $a,b,c \leq 10$~\AA{} filter used for RRUFF to yield 87{,}631 structures. Processing all six pipeline configurations against this full filtered set is computationally prohibitive, so we instead evaluated each configuration on a fixed 1{,}000-structure subset drawn uniformly at random (seed~$= 42$) from the 87{,}631-entry pool. Of the 1000-structure benchmark set, the AGAPI-XRD pipeline returned valid lattice parameters for 948 to 981 entries per workflow (Table~\ref{tab:alexandria_summary}); of the 981 structures produced without refinement, 229 (22.9\%) structurally matched the corresponding reference structures as defined by pymatgen \texttt{StructureMatcher} (\texttt{stol}~=~$0.5$). The AGAPI-XRD pipeline returned valid lattice parameters for 79.7\% of the benchmark minerals and identified a candidate structure for 93.8\%, with 17 minerals unmatched with respect to XRD pattern (Table~\ref{tab:coverage}). Pattern matching against the combined JARVIS-DFT and COD databases succeeded for 72.8\% of minerals (Figure~\ref{fig:crystal_system_histograms_plus_match_rate_pie}), while DiffractGPT independently succeeded for 81.2--83.7\%. Taken together, these two methods show substantial complementarity: pattern matching covers well-characterized stoichiometric compounds present in DFT or COD databases, whereas DiffractGPT fills gaps for complex minerals with non-standard stoichiometries, partial site occupancies, or hydrated phases. The 17 unmatched minerals share common characteristics, namely highly complex or ambiguous chemical formulas containing site-mixing notation (e.g., Aeschynite-(Y): (Y,Ca,Th)(Ti,Nb)$_2$(O,OH)$_6$) and/or rare element combinations absent from both the databases and the training data. 

\begin{table}[h]
\caption{Structure identification rate on the RRUFF benchmark (276 minerals, $a,b,c \leq 10$~\AA): the fraction of minerals for which each method returned a candidate structure. This measures identification only and does not imply correct lattice parameters; of the 276 minerals, 220 (79.7\%) yielded valid lattice parameters for the accuracy analysis (Tables~\ref{tab:overall_metrics}--\ref{tab:crystal_system}). Pattern matching uses the combined JARVIS-DFT (${\sim}76{,}000$ entries) and COD (${\sim}431{,}000$ entries) databases. DiffractGPT uses a Mistral-7B transformer fine-tuned on JARVIS-DFT structure--XRD pairs. The ``Combined'' row reflects the best result from either method. The DiffractGPT range spans the six pipeline runs.}\label{tab:coverage}
\begin{tabular}{@{}lcc@{}}
\toprule
Method & Count & Identification rate (\%) \\
\midrule
Pattern matching (formula) & 145 & 52.5 \\
Pattern matching (elements, fallback) & 56 & 20.3 \\
Pattern matching (total) & 201 & 72.8 \\
DiffractGPT & 224--231 & 81.2--83.7 \\
Combined (any method) & 259 & 93.8 \\
Unmatched & 17 & 6.2 \\
\botrule
\end{tabular}
\end{table}

Among the 220 RRUFF structures with valid lattice comparisons, the best structure was provided by pattern matching for 132 entries (60\%), by DiffractGPT for 47 entries (21\%), and by element-based pattern matching for 41 entries (19\%). Moreover, integrating COD alongside JARVIS-DFT expanded the set of minerals matched by pattern matching; several minerals, including Carrollite CuCo$_2$S$_4$ and Aegirine NaFeSi$_2$O$_6$, matched multiple COD entries alongside JARVIS-DFT entries, thereby providing independent structural validation.

The Alexandria PBE-hull subset provides a complementary simulated-XRD benchmark spanning all six pipeline configurations (Table~\ref{tab:alexandria_summary}). Across the six workflows, AGAPI-XRD returned valid lattice parameters for 948--981 entries per run, a valid lattice-parameter return rate of 94.8--98.1\%. Pattern matching alone returned a candidate for 40.2--41.6\% of entries and DiffractGPT independently for 94.3--97.6\%. Lattice-length MAEs were stable across refinement settings: $a$ remained near 0.99--1.02~\AA{}, $b$ near 1.06--1.10~\AA{}, and $c$ was larger at 1.58--1.76~\AA{}.

\begin{table}[h]
\caption{Alexandria PBE-hull simulated-XRD benchmark summary (1{,}000-structure subset, seed~$= 42$). Entries is the fixed subset size for all workflows; $N_{\mathrm{lat}}$ is the number with complete reference and predicted lattice parameters. Valid lat.\ is $N_{\mathrm{lat}}$ as a fraction of the fixed subset size, i.e.\ the valid lattice-parameter return rate.}\label{tab:alexandria_summary}
\begin{tabular}{@{}lcccccc@{}}
\toprule
Workflow & Entries & $N_{\mathrm{lat}}$ & Valid lat.\ (\%) & MAE($a$) & MAE($b$) & MAE($c$) \\
\midrule
None             & 1{,}000 & 981 & 98.1\% & 1.002 & 1.095 & 1.576 \\
None + ALIGNN-FF & 1{,}000 & 952 & 95.2\% & 0.990 & 1.079 & 1.592 \\
BGMN             & 1{,}000 & 948 & 94.8\% & 1.003 & 1.096 & 1.760 \\
BGMN + ALIGNN-FF & 1{,}000 & 972 & 97.2\% & 1.018 & 1.094 & 1.755 \\
GSAS-II          & 1{,}000 & 949 & 94.9\% & 1.001 & 1.084 & 1.603 \\
GSAS-II + ALIGNN-FF & 1{,}000 & 957 & 95.7\% & 0.989 & 1.058 & 1.600 \\
\botrule
\end{tabular}
\end{table}

\begin{figure}[t]
\centering
\makebox[\linewidth][c]{%
\includegraphics[width=1.4\linewidth]
{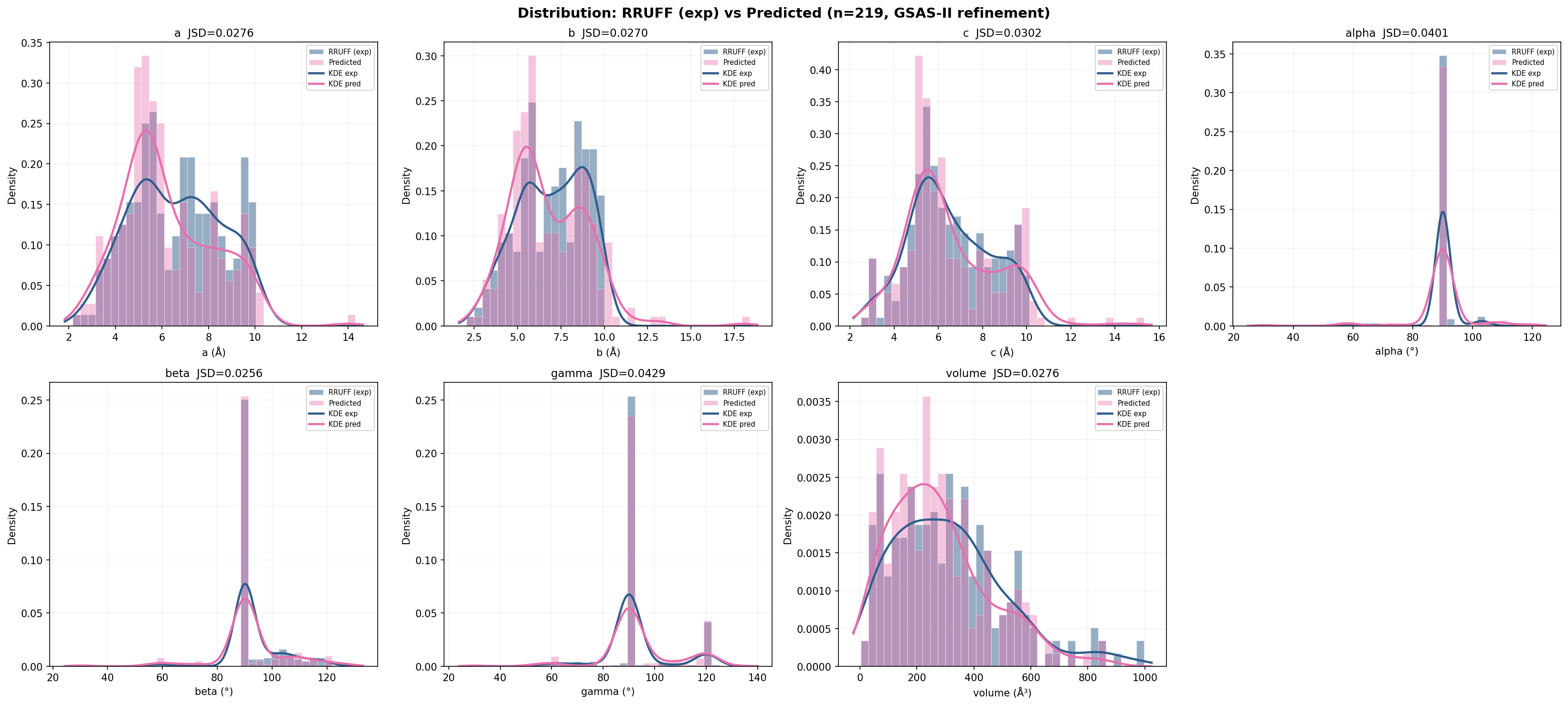}}
\caption{\textbf{Distribution analysis of actual versus predicted lattice parameters and actual versus predicted cell volumes across the RRUFF benchmark with the GSAS-II refinement method.} Histograms of cell volumes and all six lattice parameters are shown for the predicted crystal structures and the ground-truth crystal structures, with predicted corresponding to the pink distributions, ground-truth corresponding to the blue distributions, and overlap corresponding to the dark magenta regions. Kernel Density Estimate (KDE) curves corresponding to predicted and experimental histograms are shown for visualization of the histograms. For length-wise parameters, the DiffractGPT with GSAS-II refinement pipeline over-predicts smaller parameters and under-predicts larger parameters. For angular parameters, assuming that the distributions can be modeled as a sum of Gaussian curves, the pipeline seems to correctly estimate the center points of the Gaussian curves but overestimate the variance. For cell volumes, the pipeline over-predicts smaller parameters and under-predicts larger parameters, which is expected given the length-wise distributions. Moreover, Jensen--Shannon Divergence (JSD) is shown to numerically quantify the divergence between the predicted and ground-truth distributions.
}
\label{fig:error_histogram}
\end{figure}

After filtering to 220 entries with complete lattice parameter data, the overall accuracy metrics are summarized in Table~\ref{tab:overall_metrics}. In general, lattice lengths $a$, $b$, and $c$ exhibit MAEs of approximately 0.7~\AA{} with positive skill scores (+0.53 to +0.57), indicating that the pipeline meaningfully outperforms the trivial baseline of predicting the population mean. Cell volume predictions reach $R^2 = 0.45$ with a skill score of +0.56. By contrast, the low or negative $R^2$ values for the angles ($\alpha$ negative; $\beta$ and $\gamma$ near zero) do not imply random predictions. Rather, they reflect the narrow distribution of experimental angles, since most minerals have angles near $90^\circ$ or $120^\circ$, so even a modest systematic offset yields a disproportionately large $R^2$ penalty. Meanwhile, Jensen--Shannon divergence values range from 0.03 to 0.06, indicating close distributional overlap (Figure~\ref{fig:error_histogram}).

\begin{figure}[t]
\centering
\makebox[\textwidth][c]{%
  \includegraphics[width=0.7\linewidth]{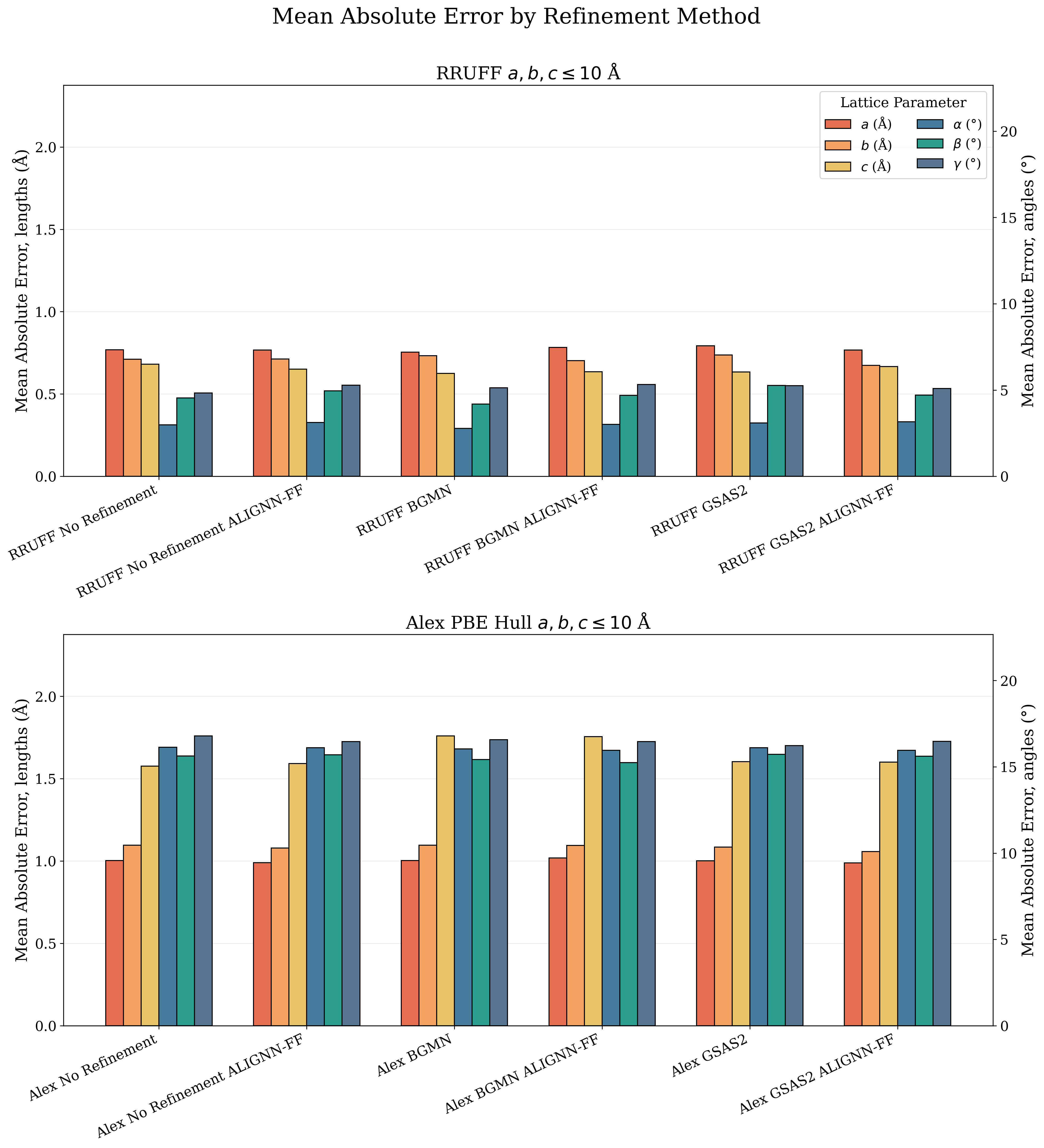}
}
\caption{\textbf{Mean Absolute Error (MAE) for six lattice parameters across RRUFF and Alexandria workflows.} Predicted AGAPI-XRD lattice parameters are compared with RRUFF experimental reference values (276 minerals) and Alexandria DFT reference values (1{,}000-structure subset, seed~$= 42$) for no refinement, BGMN, GSAS-II, and optional ALIGNN-FF relaxation on top of the chosen refinement method. Lengthwise lattice parameters are shown with warm colors, and angular lattice parameters are shown with cool colors. Vertical axes between subplots are shared within each panel to enable direct comparison. On average, RRUFF structures are predicted with lower MAE than Alexandria structures. For each benchmark, neither Rietveld refinement nor ALIGNN-FF relaxation substantially improves raw lattice reconstruction MAE.}
\label{fig:refinement_mae_all6}
\end{figure}

\begin{figure}[t]
\centering
\makebox[\textwidth][c]{%
  \includegraphics[width=0.7\linewidth]{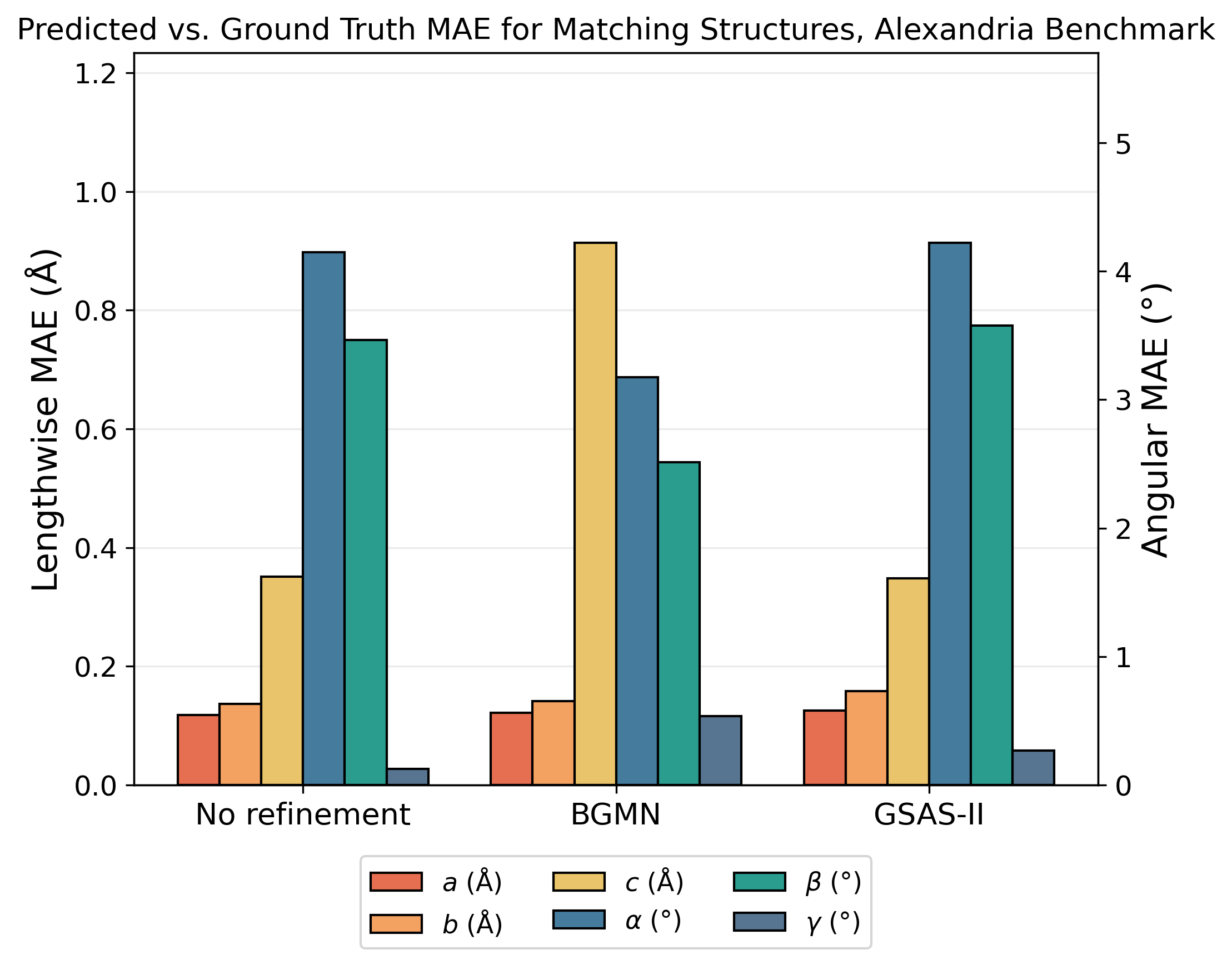}
}
\caption{\textbf{Lattice parameter MAE on StructureMatcher-matched Alexandria structures, with and without Rietveld refinement.} Predicted AGAPI-XRD lattice parameters are compared with Alexandria DFT reference values for the subset of structures whose no-refinement (vanilla DiffractGPT) prediction matches the ground truth under the \texttt{pymatgen} \texttt{StructureMatcher} at \texttt{stol}~$= 0.5$ (229 of 981 predicted structures, 22.9\% of the 1{,}000-structure subset, seed~$= 42$). The same matched identifiers are used for the no-refinement, BGMN, and GSAS-II workflows, and MAE is reported on the $n = 211$ structures with valid predictions in all three. Lengthwise parameters ($a$, $b$, $c$) are shown with warm colors on the left axis (\AA), and angular parameters ($\alpha$, $\beta$, $\gamma$) with cool colors on the right axis ($^\circ$). Restricting to structurally correct predictions, neither BGMN nor GSAS-II refinement systematically improves lattice MAE: BGMN reduces $\alpha$ and $\beta$ error but increases $c$ and $\gamma$ error, whereas GSAS-II tracks the unrefined prediction closely across all parameters.}
\label{fig:matched_refinement_mae}
\end{figure}

The aggregate MAE in Figure~\ref{fig:refinement_mae_all6} is computed over all structures with valid predictions, including those for which DiffractGPT generated a structurally incorrect cell. To isolate the effect of refinement on cases where the generative model has already recovered the correct structure, we restrict the comparison to structures whose predicted geometry matches the DFT ground truth under the pymatgen \texttt{StructureMatcher}. The match set is defined once on the no-refinement output: of the 981 structures in the Alexandria subset for which DiffractGPT produced a structure, 229 match the ground truth at this threshold. The identical set of matched identifiers is then carried over to the BGMN and GSAS-II workflows, and MAE is computed on the 211 structures with valid predictions across all three workflows (Figure~\ref{fig:matched_refinement_mae}). Even on this favorable subset, Rietveld refinement does not consistently reduce lattice MAE: BGMN improves the angular parameters $\alpha$ and $\beta$ but degrades $c$ and $\gamma$, while GSAS-II is essentially indistinguishable from the unrefined prediction across all six parameters. This reinforces the conclusion that the dominant error source is the DiffractGPT prediction itself rather than the lack of subsequent refinement. For reference, Table~\ref{tab:master_results} compiles the complete set of metrics for all twelve experimental configurations-the six RRUFF and six Alexandria workflows of Figure~\ref{fig:refinement_mae_all6}-including total and valid prediction counts, identification (match) rates, per-parameter MAE, skill scores, mean Jensen--Shannon divergence, and the within-30\% relative-accuracy rate.

\newpage

\begin{table}[p]
\centering
\caption{\textbf{Comprehensive transposed results for all twelve experimental configurations.}
Each column is one pipeline configuration. Total~=~entries processed; Valid~=~entries with complete reference and predicted lattice parameters. Identification rates are reported as percentages: Comb.~=~either pattern matching or DiffractGPT returned a candidate structure; PM~=~pattern matching (JARVIS-DFT~$+$~COD); DG~=~DiffractGPT. Lattice-parameter MAE is reported for $a$, $b$, $c$ in~\AA, $\alpha$, $\beta$, $\gamma$ in degrees, and $V$ in~\AA$^3$. Skill~$=1-\mathrm{MAE}/\mathrm{MAD}(\mathrm{exp})$ is given as the mean over the three lattice lengths ($a$--$c$) and for volume. $\overline{\mathrm{JSD}}$ is the Jensen--Shannon divergence between the predicted and reference distributions, averaged over all seven parameters. Rel.~$<\!30\%$ is the percentage of valid predictions whose mean relative length error across $a,b,c$ is below~30\%.}
\label{tab:master_results}
\footnotesize
\setlength{\tabcolsep}{4pt}

\begin{tabular}{@{}lrrrrrr@{}}
\toprule
\multicolumn{7}{@{}l}{\textit{RRUFF benchmark (276 minerals, $a,b,c \leq 10$~\AA)}}\\
\midrule
Metric
& None
& \shortstack{None\\+ ALIGNN-FF}
& BGMN
& \shortstack{BGMN\\+ ALIGNN-FF}
& GSAS-II
& \shortstack{GSAS-II\\+ ALIGNN-FF} \\
\midrule
Total & 276 & 276 & 276 & 276 & 276 & 276 \\
Valid & 220 & 217 & 216 & 217 & 219 & 216 \\
Comb. (\%) & 93.8 & 93.5 & 92.4 & 92.8 & 93.8 & 92.8 \\
PM (\%) & 72.8 & 72.8 & 72.8 & 72.8 & 72.8 & 72.8 \\
DG (\%) & 83.0 & 83.7 & 81.5 & 81.2 & 81.9 & 81.9 \\
\addlinespace
MAE $a$ & 0.77 & 0.77 & 0.75 & 0.78 & 0.79 & 0.77 \\
MAE $b$ & 0.71 & 0.71 & 0.73 & 0.70 & 0.74 & 0.67 \\
MAE $c$ & 0.68 & 0.65 & 0.63 & 0.63 & 0.63 & 0.67 \\
MAE $\alpha$ & 2.99 & 3.12 & 2.78 & 3.01 & 3.10 & 3.16 \\
MAE $\beta$ & 4.55 & 4.95 & 4.19 & 4.70 & 5.26 & 4.71 \\
MAE $\gamma$ & 4.83 & 5.29 & 5.13 & 5.33 & 5.25 & 5.08 \\
MAE $V$ & 68.8 & 68.9 & 67.6 & 69.7 & 70.8 & 67.0 \\
\addlinespace
Skill $a$--$c$ & $+$0.55 & $+$0.56 & $+$0.56 & $+$0.56 & $+$0.55 & $+$0.56 \\
Skill $V$ & $+$0.56 & $+$0.56 & $+$0.56 & $+$0.55 & $+$0.54 & $+$0.57 \\
$\overline{\mathrm{JSD}}$ & 0.034 & 0.032 & 0.031 & 0.031 & 0.032 & 0.036 \\
Rel.\,$<\!30\%$ & 92.7 & 91.7 & 93.1 & 92.6 & 91.3 & 91.2 \\
\bottomrule
\end{tabular}

\vspace{1.2em}

\begin{tabular}{@{}lrrrrrr@{}}
\toprule
\multicolumn{7}{@{}l}{\textit{Alexandria PBE-hull (1{,}000-structure subset, seed~$=42$)}}\\
\midrule
Metric
& None
& \shortstack{None\\+ ALIGNN-FF}
& BGMN
& \shortstack{BGMN\\+ ALIGNN-FF}
& GSAS-II
& \shortstack{GSAS-II\\+ ALIGNN-FF} \\
\midrule
Total & 1000 & 1000 & 1000 & 1000 & 1000 & 1000 \\
Valid & 981 & 952 & 948 & 972 & 949 & 957 \\
Comb. (\%) & 98.1 & 95.2 & 94.8 & 97.2 & 94.9 & 95.7 \\
PM (\%) & 41.6 & 40.2 & 40.5 & 41.2 & 40.3 & 40.4 \\
DG (\%) & 97.6 & 94.7 & 94.3 & 96.7 & 94.4 & 95.2 \\
\addlinespace
MAE $a$ & 1.00 & 0.99 & 1.00 & 1.02 & 1.00 & 0.99 \\
MAE $b$ & 1.10 & 1.08 & 1.10 & 1.09 & 1.08 & 1.06 \\
MAE $c$ & 1.58 & 1.59 & 1.76 & 1.76 & 1.60 & 1.60 \\
MAE $\alpha$ & 16.14 & 16.11 & 16.05 & 15.95 & 16.11 & 15.96 \\
MAE $\beta$ & 15.63 & 15.70 & 15.43 & 15.24 & 15.73 & 15.62 \\
MAE $\gamma$ & 16.79 & 16.46 & 16.57 & 16.47 & 16.23 & 16.47 \\
MAE $V$ & 38.2 & 38.5 & 43.3 & 45.0 & 38.6 & 39.3 \\
\addlinespace
Skill $a$--$c$ & $-$0.11 & $-$0.12 & $-$0.17 & $-$0.17 & $-$0.12 & $-$0.10 \\
Skill $V$ & $+$0.47 & $+$0.44 & $+$0.39 & $+$0.36 & $+$0.46 & $+$0.46 \\
$\overline{\mathrm{JSD}}$ & 0.126 & 0.127 & 0.126 & 0.125 & 0.125 & 0.126 \\
Rel.\,$<\!30\%$ & 49.8 & 50.1 & 50.6 & 50.0 & 50.5 & 50.8 \\
\bottomrule
\end{tabular}

\end{table}

\begin{table}[h]
\caption{Overall lattice parameter accuracy on the RRUFF benchmark ($N = 220$, filtered to entries with RRUFF $a,b,c \leq 10$~\AA{} and valid predictions). MAE~=~mean absolute error, RMSE~=~root mean square error, $R^2$~=~coefficient of determination, Med.~$|\Delta|$~=~median absolute error, MAD(exp)~=~mean absolute deviation of experimental values, Skill~=~$1 - \mathrm{MAE}/\mathrm{MAD}(\mathrm{exp})$, JSD~=~Jensen--Shannon divergence between experimental and predicted distributions.}\label{tab:overall_metrics}
\begin{tabular}{@{}lccccccc@{}}
\toprule
Parameter & MAE & RMSE & $R^2$ & Med.\ $|\Delta|$ & MAD(exp) & Skill & JSD \\
\midrule
$a$ (\AA) & 0.77 & 1.48 & 0.40 & 0.03 & 1.63 & +0.53 & 0.03 \\
$b$ (\AA) & 0.71 & 1.48 & 0.40 & 0.03 & 1.65 & +0.57 & 0.06 \\
$c$ (\AA) & 0.68 & 1.47 & 0.35 & 0.03 & 1.52 & +0.55 & 0.03 \\
$\alpha$ ($^\circ$) & 2.99 & 8.71 & $-$0.47 & 0.00 & 2.35 & $-$0.27 & 0.04 \\
$\beta$ ($^\circ$) & 4.55 & 10.56 & 0.05 & 0.00 & 6.88 & +0.34 & 0.03 \\
$\gamma$ ($^\circ$) & 4.83 & 12.47 & 0.12 & 0.00 & 7.76 & +0.38 & 0.04 \\
$V$ (\AA$^3$) & 68.8 & 146.2 & 0.45 & 1.9 & 154.7 & +0.56 & 0.03 \\
\botrule
\end{tabular}
\end{table}

A particularly important result is the sharp performance difference between the two identification methods, as shown in Table~\ref{tab:method_breakdown}. For pattern-matched structures, lattice length MAEs are 0.23~\AA{} to 0.33~\AA{} with consistently high skill scores of +0.80 to +0.85, and the volume skill reaches +0.81. These remaining errors likely arise from a combination of systematic DFT overestimation of lattice constants, typically 1\% to 3\% with OptB88vdW~\cite{jarvis2020}, residual cell transformation errors, and axis assignment ambiguity. DiffractGPT, in contrast, shows substantially larger MAE ranging between 1.63~\AA to 1.82~\AA with negative skill scores. Nevertheless, its primary value lies in extending structure generation to the 47 minerals that could not be identified by database search.

\begin{table}[h]
\caption{Lattice parameter MAE and Skill stratified by identification method. Pattern matching (PM) draws structures from the JARVIS-DFT and COD databases. DiffractGPT (DG) generates structures de novo from XRD peaks and formula. PM~(elements) is the fallback compositional search triggered when exact formula matching fails. Skill values above zero indicate the method outperforms predicting the experimental population mean.}\label{tab:method_breakdown}
\begin{tabular}{@{}lcccccc@{}}
\toprule
& \multicolumn{2}{c}{PM ($n=132$)} & \multicolumn{2}{c}{DG ($n=47$)} & \multicolumn{2}{c}{PM elem.\ ($n=41$)} \\
\cmidrule(lr){2-3}\cmidrule(lr){4-5}\cmidrule(lr){6-7}
Param.\ & MAE & Skill & MAE & Skill & MAE & Skill \\
\midrule
$a$ (\AA) & 0.33 & +0.80 & 1.64 & $-$0.31 & 1.18 & +0.21 \\
$b$ (\AA) & 0.26 & +0.85 & 1.63 & $-$0.30 & 1.11 & +0.24 \\
$c$ (\AA) & 0.23 & +0.84 & 1.82 & $-$0.35 & 0.83 & +0.38 \\
$V$ (\AA$^3$) & 28.4 & +0.81 & 154.2 & $-$0.10 & 100.9 & +0.17 \\
\botrule
\end{tabular}
\end{table}

Performance also varies by crystal system (Table~\ref{tab:crystal_system}). Most systems perform comparably well, with lattice-length skill scores between +0.54 and +0.65. Orthorhombic, tetragonal, and rhombohedral systems are the strongest, although the rhombohedral estimate rests on only seven minerals. Triclinic systems are the clear exception, showing negative skill for both length and volume, which is consistent with the greater difficulty of recovering structures that contain six independent lattice parameters from one-dimensional powder diffraction data.

\begin{table}[h]
\caption{Lattice parameter accuracy stratified by RRUFF crystal system classification. $n$~=~number of minerals in each system within the filtered benchmark. MAE and Skill are shown for lattice parameter $a$ (representative of lengths) and unit cell volume $V$. Most systems achieve comparable length accuracy, while triclinic systems show the poorest, consistent with the number of independent lattice parameters that must be determined from one-dimensional powder diffraction data.}\label{tab:crystal_system}
\begin{tabular}{@{}lccccc@{}}
\toprule
Crystal System & $n$ & MAE($a$) (\AA) & Skill($a$) & MAE($V$) (\AA$^3$) & Skill($V$) \\
\midrule
Cubic & 50 & 0.78 & +0.54 & 84.4 & +0.66 \\
Orthorhombic & 45 & 0.56 & +0.65 & 39.5 & +0.64 \\
Monoclinic & 45 & 0.72 & +0.58 & 54.2 & +0.43 \\
Hexagonal & 26 & 0.90 & +0.55 & 78.5 & +0.57 \\
Tetragonal & 21 & 0.54 & +0.62 & 41.0 & +0.69 \\
Triclinic & 26 & 1.32 & $-$0.54 & 140.4 & $-$0.40 \\
Rhombohedral & 7 & 0.49 & +0.65 & 21.1 & +0.88 \\
\botrule
\end{tabular}
\end{table}

% The elemental composition analysis further supports the view that lattice recovery is governed primarily by geometric rather than chemical constraints. Specifically, the distribution of prediction errors across the periodic table (Figure~\ref{fig:periodic_errors}) is broadly uniform across most elements. Where elevated errors do appear, they are concentrated mainly among sparsely represented elements, suggesting that the dominant limitation is data scarcity rather than chemistry-specific failure modes.

% Finally, the distributions of relative prediction errors for $a$, $b$, and $c$ (Figure~\ref{fig:error_histogram}) exhibit a characteristic bimodal structure, with one peak at 0--10\% relative error corresponding primarily to pattern-matched structures and a second peak at 30--40\% corresponding largely to DiffractGPT structures. Consistent with this interpretation, approximately 40\% of predictions achieve less than 10\% relative error, while 65\% fall within 30\% (Figure~\ref{fig:cumulative_errors}).

The most significant finding is the strong complementarity between database-driven pattern matching and generative AI. In particular, pattern matching provides high-accuracy identification for the 73\% of minerals with matching database entries, with skill scores of $+0.8$ to $+0.85$ for lattice lengths. Further, DiffractGPT extends identification to 94\% by generating approximate structures for complex minerals. Thus, this two-tier architecture, in which reliable database search is prioritized and AI generation serves as a fallback, maximizes both accuracy and the fraction of minerals for which a structure is returned. In addition, the element-based fallback provides an intermediate tier, successfully identifying 56 additional minerals.

At the same time, several systematic error sources contribute to the observed MAEs. First, systematic bias arises from the choice of exchange-correlation functional in the density functional theory calculations. Specifically, the JARVIS-DFT database uses the OptB88vdW functional~\cite{jarvis2020}, which systematically overestimates lattice constants by 1\% to 3\%. Second, cell setting mismatch remains an important source of error. Furthermore, the ambiguity of the inverse problem is fundamental to the task itself. Because powder XRD data is inherently one-dimensional, the mapping from diffraction pattern to crystal structure is not one-to-one~\cite{patterson1944}. 

Moreover, across three RRUFF refinement configurations, the overall metrics showed only modest variation (Figure~\ref{fig:refinement_mae_all6}). The Alexandria subset benchmark showed the same trend. Thus, for pattern-matched structures that are already close to the reference, refinement provides only marginal improvement. By contrast, for DiffractGPT structures with incorrect symmetry, refinement tends to converge to local $R_{\mathrm{wp}}$ minima. Therefore, the primary value of refinement lies less in dramatically improving lattice accuracy and more in providing quantitative figures of merit, as well as enabling more fine-grained extraction of atomic positions and thermal parameters.

More broadly, the AGAPI-XRD pipeline provides an end-to-end automated benchmark of combined database search and generative AI for crystal structure identification from powder XRD, evaluated against a large experimental mineral database. By comparison, previous approaches focused on individual components~\cite{park2017, lee2020xrd, vecsei2019}. Likewise, DiffractGPT~\cite{diffractgpt2025} was evaluated primarily on simulated data. Accordingly, our benchmark complements these efforts with a quantitative assessment against experimentally measured patterns.

Nevertheless, several limitations remain. The RRUFF benchmark is biased toward common minerals, so performance on novel synthetic materials may differ. The Alexandria benchmark broadens the chemical and structural scale of the evaluation, but it is based on simulated XRD profiles and DFT-relaxed reference structures rather than experimental measurements. Similarly, the cell-matching protocol does not guarantee correct settings for all space groups, and DiffractGPT was trained exclusively on JARVIS-DFT data. Looking ahead, several directions could improve the framework further, such as fine-tuning DiffractGPT on lower-symmetry structures, jointly optimizing cell-matching and pattern similarity scores, active learning from refinement results, and extension to multiphase identification and amorphous quantification.

%% =========================================================================
%%  CONCLUSION
%% =========================================================================
\section{Conclusion}\label{sec:conclusion}

We have presented AGAPI-XRD, a hybrid framework that combines database-driven pattern matching, DiffractGPT generative structure prediction, and automated Rietveld refinement for X-ray diffraction analysis. Benchmarked on 276 minerals from the experimental RRUFF database, the pipeline returned valid lattice parameters for 79.7\% of the minerals and identified a candidate structure for 93.8\% of the minerals. Lattice-parameter mean absolute errors were approximately 0.7~\AA{} for unit-cell lengths, and skill scores were positive for lattice lengths and volume. Across six configurations on the 1000-structure Alexandria PBE-hull simulated-XRD benchmark, the combined workflow achieved a 94.8\% to 98.1\% valid lattice-parameter return rate with lattice-length MAEs of 0.99~\AA{} to 1.76~\AA{}. The results show that the strongest performance comes from the complementarity of the framework components: pattern matching provides the highest structural accuracy when suitable database entries exist, while DiffractGPT substantially expands structure generation to complex minerals and DFT reference structures that are absent from existing reference sets. Across RRUFF crystal systems, performance is strong across most crystal systems and weakest for triclinic materials, reflecting the greater ambiguity of recovering low-symmetry structures from one-dimensional powder diffraction data. Although automated refinement provides only modest improvements in aggregate lattice metrics, it remains valuable for improving structural consistency and supplying quantitative figures of merit for candidate solutions. Taken together, these results establish AGAPI-XRD as a practical and accessible step toward automated, end-to-end crystal structure determination from powder XRD data. The pipeline is freely accessible at \url{https://atomgpt.org/xrd}.

%% =========================================================================
%%  METHODS
%% =========================================================================
\section{Methods}\label{sec:methods}

DiffractGPT is a generative pretrained transformer (GPT) that predicts crystal
structures directly from powder XRD peak data. Following the architecture of
decoder-only transformers~\cite{vaswani2017attention}, the model processes an
input sequence of tokenized peak positions and relative intensities alongside a
chemical formula token. The core operation is scaled dot-product self-attention,
\begin{equation}
  \mathrm{Attention}(Q,K,V) =
    \mathrm{softmax}\!\left(\frac{QK^{\!\top}}{\sqrt{d_k}}\right)V,
\end{equation}
where $Q$, $K$, and $V$ are learned linear projections of the input embeddings
and $d_k$ is the key dimension. Multi-head attention concatenates $h$ such heads
to capture dependencies across different representation subspaces. A fine-tuned
Mistral-7B~\cite{jiang2023mistral,diffractgpt2025} backbone maps this peak-and-formula sequence
to a structured output containing lattice parameters, space group, and fractional
atomic coordinates.

To complement generative prediction, ALIGNN-FF~\cite{choudhary2023unified}
provides optional pre-refinement structural relaxation. ALIGNN constructs two
message-passing graphs from a crystal structure: an atomic graph $\mathcal{G}$
with nodes representing atoms and edges representing bonds within a cutoff
radius, and a line graph $\mathcal{L}(\mathcal{G})$ whose nodes correspond to
bonds and whose edges encode bond angles. Node and edge features are updated by
interleaved message-passing over both graphs,
\begin{equation}
  \mathbf{h}_i^{(l+1)} = \phi\!\left(\mathbf{h}_i^{(l)},\,
    \sum_{j \in \mathcal{N}(i)} \psi\!\left(\mathbf{h}_i^{(l)},
      \mathbf{e}_{ij}^{(l)}, \mathbf{h}_j^{(l)}\right)\right),
\end{equation}
where $\mathbf{h}_i^{(l)}$ and $\mathbf{e}_{ij}^{(l)}$ are node and edge
embeddings at layer $l$, and $\phi$, $\psi$ are learnable functions. ALIGNN-FF
was trained on approximately 4.9~million DFT energy and force evaluations from
the JARVIS-DFT database and is used here to reduce geometric artifacts in
AI-generated structures prior to Rietveld refinement.

Two classical refinement engines are supported in the pipeline. GSAS-II~\cite{toby2013}
is an open-source, full-featured crystallographic software suite that implements
full-pattern Rietveld refinement via a Python scripting interface; it is widely
used in the community and provides comprehensive control over background
modeling, peak profiles, and structural parameters. BGMN~\cite{bgmn1998} is an
alternative Rietveld engine accessed through the \texttt{dara.refine} package,
designed for fast, automated multi-phase refinement; its inclusion enables
cross-validation of refinement results between two independent implementations.

The pattern matching stage simulates theoretical powder XRD patterns for each
candidate structure in the JARVIS-DFT and COD databases using the kinematic
diffraction approximation implemented in the JARVIS analysis module~\cite{jarvis2020}.
The similarity between experimental and calculated patterns is quantified by
cosine similarity computed on re-binned intensity profiles,
\begin{equation}
S = \frac{\mathbf{y}_{\mathrm{exp}} \cdot \mathbf{y}_{\mathrm{calc}}}
         {\|\mathbf{y}_{\mathrm{exp}}\| \; \|\mathbf{y}_{\mathrm{calc}}\|}
\label{eq:cosine}
\end{equation}
where $\mathbf{y}_{\mathrm{exp}}$ and $\mathbf{y}_{\mathrm{calc}}$ are the
normalized intensity vectors re-binned onto a common $2\theta$ grid with
$0.1^\circ$ spacing over the range $0$--$90^\circ$.

The RRUFF Project~\cite{rruff2015} provides the experimental benchmark used here.
RRUFF powder patterns are collected on natural mineral specimens under standardized
conditions using Cu~K$\alpha$ radiation ($\lambda = 1.54184$~\AA), and each
entry reports a continuous intensity profile spanning approximately
$20^\circ$--$95^\circ$ $2\theta$ along with experimentally determined lattice
parameters and the mineral's chemical formula; crucially, atomic coordinates are
generally not provided, so the benchmark tests lattice-level rather than
full-structural agreement.

The Alexandria PBE-hull dataset~\cite{alexandria} comprises approximately
5~million crystal structures optimized at the PBE level of DFT across the full
compositional space. For the prototype database and simulated benchmark used in
this work, only structures with convex-hull energy $E_{\mathrm{hull}} = 0$~eV/atom
are retained, on the grounds that thermodynamically stable compositions are the
most likely to be encountered experimentally. This hull-filtering reduces the
Alexandria contribution to roughly 117,000 entries while ensuring that structures
used as substitution templates and simulated-XRD references correspond to
realistically synthesizable phases. For the Alexandria benchmark, the DFT-relaxed
lattice parameters serve as the reference values and the XRD patterns are
simulated on the same fixed grid used by the AGAPI-XRD request pipeline.

The AGAPI platform exposes XRD analysis through a RESTful endpoint
(\texttt{POST /xrd/analyze\_with\_refinement}) that accepts a JSON payload
containing the digitized intensity profile, peak list, and chemical formula or
elemental composition. Optional request fields select the pattern-matching
database (JARVIS-DFT, COD, or both), enable DiffractGPT generation, choose the
Rietveld engine (GSAS-II or BGMN), activate ALIGNN-FF relaxation, and request
downstream property predictions from ALIGNN and SlaKoNet. The endpoint returns
a ranked list of candidate structures as POSCAR files together with cosine
similarity scores, $R_{\mathrm{wp}}$ values where refinement was performed,
and any requested property predictions.

Comparing predicted and reference lattice parameters requires careful treatment
of cell-setting conventions. DFT databases typically store primitive cells,
whereas RRUFF reports conventional cells. To account for this discrepancy, we
implemented an automated cell-matching protocol.

For each predicted structure, three cell variants are generated: the original
cell, the primitive cell obtained via space-group analysis in JARVIS, and the
conventional cell. For each of these variants, all six permutations of $(a,b,c)$
are evaluated with the corresponding reassignment of lattice angles. In standard
crystallographic notation, $\alpha=\angle(\mathbf{b},\mathbf{c})$,
$\beta=\angle(\mathbf{a},\mathbf{c})$, and $\gamma=\angle(\mathbf{a},\mathbf{b})$,
so any permutation of the lattice vectors requires the angles to be remapped
consistently. The optimal cell variant and axis permutation are then selected by
minimizing a composite score that combines relative length errors, relative angle
errors, and volume error. In total, this procedure evaluates $3 \times 6 = 18$
candidate cell representations per structure, substantially reducing systematic
errors arising from cell-setting mismatches.

Lattice parameter accuracy is assessed using several complementary metrics, since
no single quantity fully captures agreement between predicted and experimental
structures. The mean absolute error measures the average magnitude of the
deviation between prediction and experiment,
\begin{equation}
\mathrm{MAE} = \frac{1}{N}\sum_{i=1}^{N}\left|y_i^{\mathrm{pred}} - y_i^{\mathrm{exp}}\right|.
\end{equation}
Because the absolute value weights all deviations linearly, MAE provides a direct
and interpretable measure of the typical prediction error in the original units
of the lattice parameter.

To complement this, we also report the root mean square error,
\begin{equation}
\mathrm{RMSE} = \sqrt{\frac{1}{N}\sum_{i=1}^{N}\left(y_i^{\mathrm{pred}} - y_i^{\mathrm{exp}}\right)^2}.
\end{equation}
Unlike MAE, RMSE places greater emphasis on larger deviations because the errors
are squared before averaging. As a result, it is more sensitive to outliers and
is useful for identifying cases in which a small number of poor predictions
disproportionately affect performance.

We further evaluate agreement using the coefficient of determination,
\begin{equation}
R^2 = 1 - \frac{\sum\left(y_i^{\mathrm{pred}} - y_i^{\mathrm{exp}}\right)^2}
                 {\sum\left(y_i^{\mathrm{exp}} - \bar{y}^{\mathrm{exp}}\right)^2}.
\end{equation}
This quantity measures how well the predicted values reproduce the variance of
the experimental data. Values of $R^2$ closer to 1 indicate that the predictions
more faithfully track the experimental trends.

Because absolute error metrics alone do not indicate whether the model improves
upon a trivial reference, we also compute a skill score,
\begin{equation}
\mathrm{Skill} = 1 - \frac{\mathrm{MAE}}{\mathrm{MAD}(\mathbf{y}^{\mathrm{exp}})},
\end{equation}
where
\begin{equation}
\mathrm{MAD} = \frac{1}{n}\sum_{i=1}^{n} \left| y_i - \bar{y} \right|
\end{equation}
is the mean absolute deviation of the experimental values. This score compares
the model against a baseline that always predicts the mean of the experimental
distribution. A skill score greater than zero indicates that the model outperforms
this baseline, while a value of 1 corresponds to perfect predictions.

In addition to pointwise metrics, we assess distribution-level agreement using
the Jensen--Shannon divergence. This is useful because two models can have similar
average errors while still producing noticeably different parameter histograms.
Let $P(i)$ and $Q(i)$ denote the normalized histogram probabilities of the
predicted and experimental values in bin $i$, and let
\begin{equation}
M(i)=\frac{1}{2}\bigl(P(i)+Q(i)\bigr)
\end{equation}
be the midpoint distribution. The Jensen--Shannon divergence is then defined as
\begin{equation}
\mathrm{JSD}(P \,\|\, Q)
=
\frac{1}{2} D_{\mathrm{KL}}(P \,\|\, M)
+
\frac{1}{2} D_{\mathrm{KL}}(Q \,\|\, M),
\end{equation}
where the Kullback--Leibler divergence is
\begin{equation}
D_{\mathrm{KL}}(P \,\|\, Q)=\sum_i P(i)\log\!\left(\frac{P(i)}{Q(i)}\right).
\end{equation}
Smaller values of JSD indicate closer agreement between the predicted and
experimental distributions, with $\mathrm{JSD}=0$ corresponding to identical
distributions.

To avoid numerical issues from zero-probability bins, we apply Laplace smoothing
before normalization,
\begin{equation}
P(i)=\frac{n_i+\alpha}{\sum_{j=1}^{K}(n_j+\alpha)},
\end{equation}
with an analogous expression for $Q(i)$, where $n_i$ is the count in bin $i$,
$K$ is the number of bins, and $\alpha>0$ is a small constant. This ensures that
every bin has nonzero support and allows the divergence to be computed in a
stable and consistent way.

%% =========================================================================
%%  DECLARATIONS
%% =========================================================================
\section*{Data Availability}

All code, benchmark scripts, and analysis notebooks are available at \url{https://github.com/atomgptlab/agapi_xrd_paper}. The RRUFF powder XRD data is openly available from the RRUFF Project~\cite{rruff2015} at \url{https://rruff.info} (bulk powder-diffraction downloads at \url{https://www.rruff.net/zipped_data_files/powder/}), and was accessed in this work through the JARVIS figshare interface. The 1000-structure subset of the Alexandria PBE-hull can be obtained via \url{https://github.com/atomgptlab/agapi_xrd_paper/blob/main/datasets/alex_subset_1000.json}. The AGAPI-XRD API is publicly available at \url{https://atomgpt.org/xrd}.

\section*{Acknowledgements}

K.C.\ acknowledges support from the Johns Hopkins University Discovery Award and the National Institute of Standards and Technology.

\section*{Author Contributions}

K.C.\ conceived and supervised the project. J.E.\ and C.R.C.\ implemented the benchmark pipeline and analysis. J.L.\ contributed to DiffractGPT model integration. F.M.A.\ contributed to Rietveld refinement methodology. All authors discussed results and contributed to manuscript preparation.

\section*{Competing Interests}

The authors declare no competing interests.

\section*{Disclaimer}

The views expressed in this article are those of the authors and do not reflect the official policy or position of the U.S. Naval Academy, Department of the Navy, the Department of War, or the U.S. Government.

\bibliography{references}
\bibliographystyle{sn-mathphys-num}

\end{document}